\documentclass[fleqn,usenatbib,useAMS]{mnras}

\usepackage{graphicx}	
\usepackage{amsmath}	
\usepackage{amssymb}	
\usepackage{multicol}        
\usepackage{bm}		
\usepackage{pdflscape}	
\usepackage{xcolor}	

\usepackage[T1]{fontenc}
\usepackage{ae,aecompl}

\usepackage{newtxtext,newtxmath}

\title[Accretion Changes in a Polluted White Dwarf]{Accretion Rate Changes Detected in a Polluted White Dwarf}

\author[J.~Farihi et al.]{Jay Farihi$^1$\thanks{E-mail: j.farihi@ucl.ac.uk},
Hiba Tu Noor$^1$,
Carl Melis$^2$,
Beth L.~Klein$^3$,
Snehalata Sahu$^4$,
Boris T.~G\"ansicke$^4$,
\newauthor
Mark C.~Wyatt$^5$,
Seth Redfield$^6$,
and
Ted M.~Johnson$^7$
\\
$^1$Department of Physics and Astronomy, University College London, London WC1E 6BT, UK\\
$^2$Department of Astronomy \& Astrophysics, University of California San Diego, La Jolla CA 92093-0424, USA\\
$^3$Department of Physics and Astronomy, University of California, Los Angeles, CA 90095-1562, USA\\
$^4$Department of Physics, University of Warwick, Coventry CV4 7AL, UK\\
$^5$Institute of Astronomy, University of Cambridge, Cambridge CB3 0HA, UK\\
$^6$Astronomy Department and Van Vleck Observatory, Wesleyan University, Middletown, CT 06459, USA\\
$^7$Department of Physics and Astronomy, University of Nevada, Las Vegas, NV 89154, USA}

\begin{document}

\maketitle

\begin{abstract}
This letter reports statistically significant changes in the equivalent widths of \ion{Mg}{ii} and \ion{Ca}{ii} lines in the dusty and polluted white dwarf WD\,0106$-$328, based on six epochs of spectroscopy using the VLT and Keck spanning 25\,yr.  Furthermore, the ratio of these two equivalent widths may also vary, with a 7\,per cent probability of being constant.  Between 2000 and 2025, both Mg and Ca have experienced decreases in accretion rates, of approximately 20 and 60\,per cent, respectively, but with individual variation during the interim.  These metal abundance decreases are the first empirical corroboration of diffusion theory in white dwarfs, which predict sinking timescales on the order of days for this star.  However, the persistent atmospheric metals require a more gradual, circumstellar process, where one possibility is viscous spreading in an ionized disk of metals, consistent with $\upalpha\approx0.1$ within that formalism. The combination of optical and ultraviolet spectroscopy with the {\em Hubble Space Telescope} detects all the major rock-forming elements (O, Mg, Si, Fe), and demonstrates that Fe dominates the accreted material by mass, and that it is delivered mostly as pure metal from within a differentiated parent body.  This inference is consistent with the possibility that chemically-segregated accretion may result from a combination of planetary assembly, fragmentation, disk evolution, and be observed on relatively short timescales.
\end{abstract}


\section{Introduction}

Metals in white dwarf atmospheres carry one of the most powerful observational signatures of exoplanetary composition and assembly.  The high surface gravities of white dwarfs should result in the rapid removal of all elements heavier than hydrogen or helium \citep[e.g.][]{koester2009a,cunningham2019}, but instead white dwarfs are commonly found to be externally polluted by metals \citep{zuckerman2010,ouldrouis2024}.  The deposition of heavy elements must be ongoing in the majority of cases, where the necessary circumstellar reservoirs of dust and gas are often detected via their emission \citep{farihi2016a,manser2020}, and now in over a dozen cases via transits of their disk material \citep{guidry2021,bhattacharjee2025}

The planetary material accreted by white dwarfs has been modeled as star-grazing planetesimals that undergo catastrophic fragmentation via tidal forces and collisions, and eventually form relatively compact disks that supply heavy elements to the stellar surface \citep[e.g.][]{jura2003,brouwers2022}.  The metal-enriched photospheres of white dwarfs reveal the bulk chemical composition of the disrupted and accreted parent body, and provide a unique window into the leftover building blocks or fragments of exoplanets or possibly their moons \citep[e.g.][]{jura2014,doyle2021}.  To date, some key insights revealed using white dwarf pollution include: evidence for differentiated parent bodies \citep{gansicke2012}, the presence of substantial water in otherwise-rocky bodies \citep{farihi2013c}, and widespread Earth-like geochemistry \citep{doyle2019}.

\begin{table*}
\begin{center}
\caption{Multi-epoch equivalent width measurements and abundance determinations using \ion{Ca}{ii}\,K and \ion{Mg}{ii}\,4482\AA \label{obs}.}
\begin{tabular}{lccccrccccc}
		
\hline

UT Date		&Facility	&Instrument	&Program	&PI        	&Resolving	&S/N  	&$W_\ion{Mg}{ii}$	&$W_\ion{Ca}{ii}$ 	&[Mg/H]   	&[Ca/H]\\
			&		&			&             	&          	&power		&(4500\,\AA)&(m\AA)	      	&(m\AA)			&(dex)	&(dex)\\

\hline

2000.12.07	&VLT	&UVES	&165.H-0588  	&Napiwotzki 	&20\,000	&14		&$104\pm22$ 	&$107\pm9$	&$-5.69\pm0.12$     &$-6.13\pm0.10$\\
2001.08.25	&VLT	&UVES	&167.D-0407	&Napiwotzki 	&20\,000	&15		&$109\pm15$ 	&$111\pm8$	&$-5.67\pm0.08$     &$-6.09\pm0.08$\\
2008.08.07	&Keck	&HIRES	&U062Hb      	&Jura		 &37\,000	&53		&$106\pm4$    	&$88\pm3$	&$-5.68\pm0.02$     &$-6.33\pm0.03$\\
2018.11.07	&VLT	&X-shooter&101.C-0646  	&G\"ansicke 	&4100	&220		&$114\pm8$    	&$99\pm5$	&$-5.64\pm0.04$     &$-6.22\pm0.05$\\
2022.06.12	&VLT	&UVES 	&109.23EK    	&Farihi     		&20\,000	&17		&$120\pm10$   &$82\pm4$	&$-5.61\pm0.05$     &$-6.45\pm0.06$\\
2025.10.19	&Keck	&HIRES	&N078        	&Redfield	 	&37\,000	&110		&$87\pm3$     	&$79\pm2$	&$-5.78\pm0.02$     &$-6.46\pm0.03$\\

\hline

\end{tabular}
\end{center}
\end{table*}

The physics underlying the link between the observed photospheric abundances and those in the accreted parent body or bodies is atomic diffusion theory.  Its application to white dwarfs has been understood qualitatively for nearly half a century \citep{fontaine1979,vauclair1979b}, with quantitative modeling beginning four decades ago \citep{paquette1986}.  As a typical white dwarf cools below below $T_{\rm eff}\approx25\,000$\,K, radiative forces become negligible compared to downward diffusion, resulting in metal-free atmospheres on timescales that are always orders of magnitude shorter than their cooling ages, and in excellent agreement with numerous, apparently pristine hydrogen and helium atmospheres, even at high spectral resolution \citep{zuckerman2003,koester2005a}.  

The heavy element sinking timescales generally scale with the size of the convection zone, and are modeled to reach $10^5$\,yr and $10^7$\,yr in the coolest known hydrogen and helium-rich white dwarfs of typical mass, respectively \citep{koester2020}.  In contrast, it has long been recognized that in warmer, hydrogen-rich white dwarfs, which lack a sizable convection zone (a typical, fully mixed layer is $10^{-16}$\,M$_\odot$), atmospheric metals should diffuse below the photosphere on the order of days.  Despite this theoretical expectation, actual data are sparse.  In the prototype dusty white dwarf G29-38, there has been a claim that the \ion{Ca}{ii}\,K line is variable \citep{vonhippel2007}, but this was subsequently called into question based on an independent study, citing the ZZ\,Ceti (variable) nature of this star \citep{debes2008}.  Only two other polluted white dwarfs with predicted short sinking timescales have published, multi-epoch observations of their metal line strengths; both GD\,56 and G238-44 exhibit tightly correlated equivalent widths of their \ion{Ca}{ii}\,K and \ion{Mg}{ii}\,4482\AA lines over 8--10 epochs, and with one potential outlying epoch each that cannot be confirmed as real \citep{farihi2018c,johnson2022}.

This work reports variations in the metal absorption lines of the white dwarf WD\,0106$-$328.  These changes are observed in both \ion{Ca}{ii}\,K and \ion{Mg}{ii}\,4482\AA, using two independent instruments, each with at least two epochs.  Section 2 describes the various observational data sets, Section 3 details the measurements and results, followed by a discussion and summary in Sections 4 and 5.

\section{Observations and data}

WD\,0106$-$328 (= HE\,0106$-$3253, Ton\,S\,193, hereafter 0106$-$328; $T_{\rm eff}=16\,100$\,K, $\log\,g=8.01$, see Section 3) was observed as part of a larger program (P109.23EK) to examine potential long-term variability in polluted white dwarfs, where all targets were chosen to be hydrogen-rich stars of DAZ spectral type and diffusion timescales less than 1\,yr (Noor et al.\ in preparation).  Spectroscopy was carried out on 2022 Jun 12 using the Ultraviolet and Visual Echelle Spectrograph \citep[UVES;][]{dekker2000} on the European Southern Observatory (ESO) Very Large Telescope (VLT).  The instrument was configured with a standard dichroic with central wavelengths 3900 and 5640\,\AA, a 2.1\,arcsec slit, and arrays with $2\times2$ binning, and slow readout.  Two exposures of 915\,s were taken in clear weather conditions with average seeing better than 1.6\,arcsec.

This setup yields a nominal resolving power of $R\approx20\,000$ and was chosen to precisely match the same instrument configuration used during the Supernova Progenitor Survey \citep[SPY;][]{napiwotzki2003} two decades prior, which observed numerous white dwarfs and detected metal lines for the first time in 18 hydrogen-rich stars, including 0106$-$328 \citep{koester2005a}.  The 2022 observations are complemented by these archival spectra obtained as part of SPY, taken on 2000 Dec 7 and 2001 Aug 25 using identical instrument settings, each with a total of two exposures of 600\,s each.

For all three UVES epochs, the raw science and calibration frames were retrieved from the ESO archive and reduced using the UVES workflow in \textsc{reflex}. Each spectrum was bias-subtracted, flat-fielded, wavelength-calibrated, and extracted according to the recommended procedures in the \textsc{reflex} UVES documentation.  Each pair of consecutive exposures were combined by a weighted average to improve signal-to-noise (S/N), and subsequently normalized. 

The target was also observed with X-shooter \citep{vernet2011} on the VLT on 2018 Nov 7.  Observations were conducted in staring mode using a 1.3\,arcsec slit in the UVB arm, yielding a resolving power $R\approx4100$ over the nominal wavelength range 3000--5560\,\AA.  The data consisted of four consecutive 900\,s exposures, and were reduced using the X-shooter workflow within \textsc{reflex}, following observatory guidelines. The individual spectra were then co-added and normalized.

Spectroscopy was carried out twice using the High-Resolution Echelle Spectrometer \citep[HIRES;][]{vogt1994} on Keck I, on both 2008 Aug 07 and 2025 Oct 19.  The setup was the same each time, with the C5 decker and a 1.15\,arcsec slit, and with the blue cross-disperser resulting in a wavelength coverage of 3100--5900\,\AA.  Both datasets, including flat field and arc lamp calibration frames, were processed with {\sc makee} and individual exposures were co-added into a single spectrum for subsequent measurements.

Ultraviolet spectroscopy using the Cosmic Origins Spectrograph \citep[COS;][]{green2012} on the {\em Hubble Space Telescope (HST)} was carried out on 28 Aug 2016 (Program 14597, PI Farihi).  The instrument was set up with the G130M grating, which has a nominal resolving power $R\approx16\,000$ using a central wavelength of 1291\,\AA, and covering (with a small gap) the range 1130–1435\,\AA.  The star was acquired and moved onto the spectrographic slit, then observed over a single orbit, using one each of all four FP-POS positions per orbit, for a total exposure time of 2215\,s.   The spectra were processed with the {\sc calcos} pipeline v3.1.8 which produced flux-calibrated spectra with S/N $\approx10$ in the region 1320--1330\,\AA.

\begin{figure}
\includegraphics[width=\columnwidth]{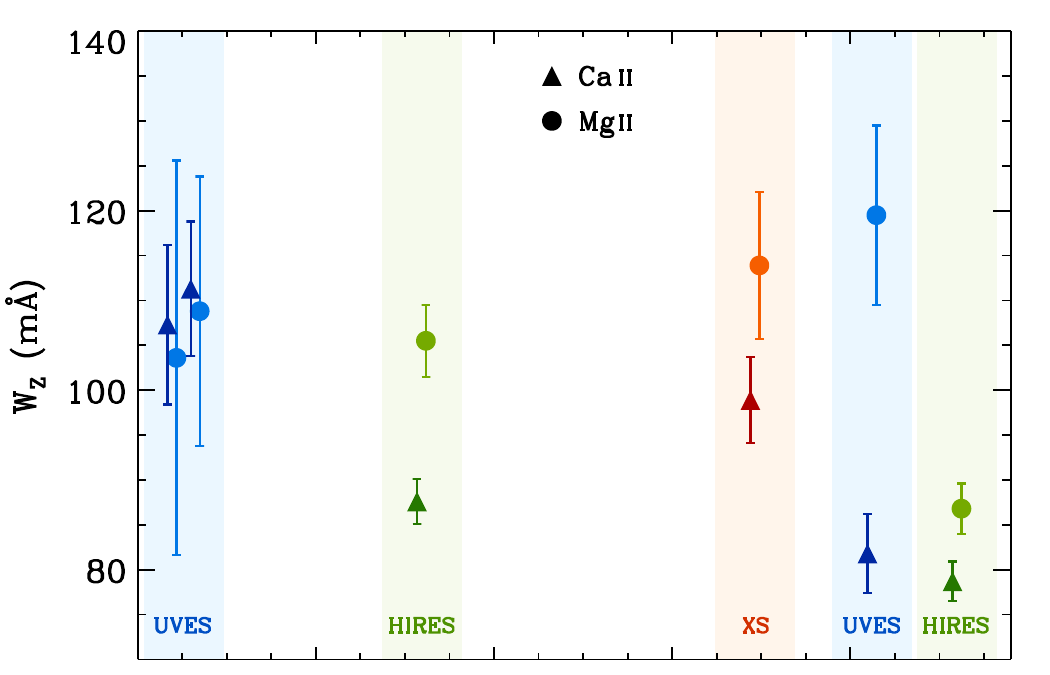}
\includegraphics[width=\columnwidth]{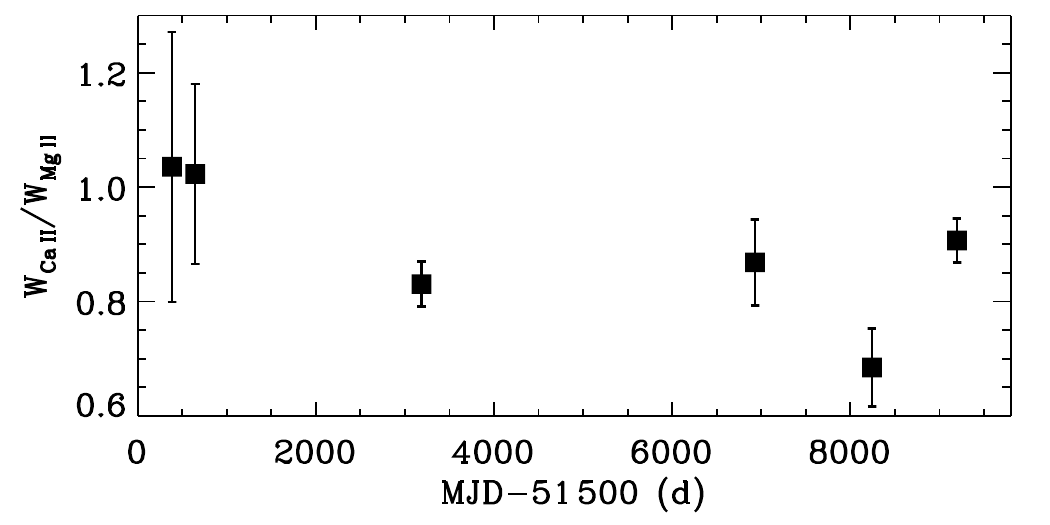}
\includegraphics[width=\columnwidth]{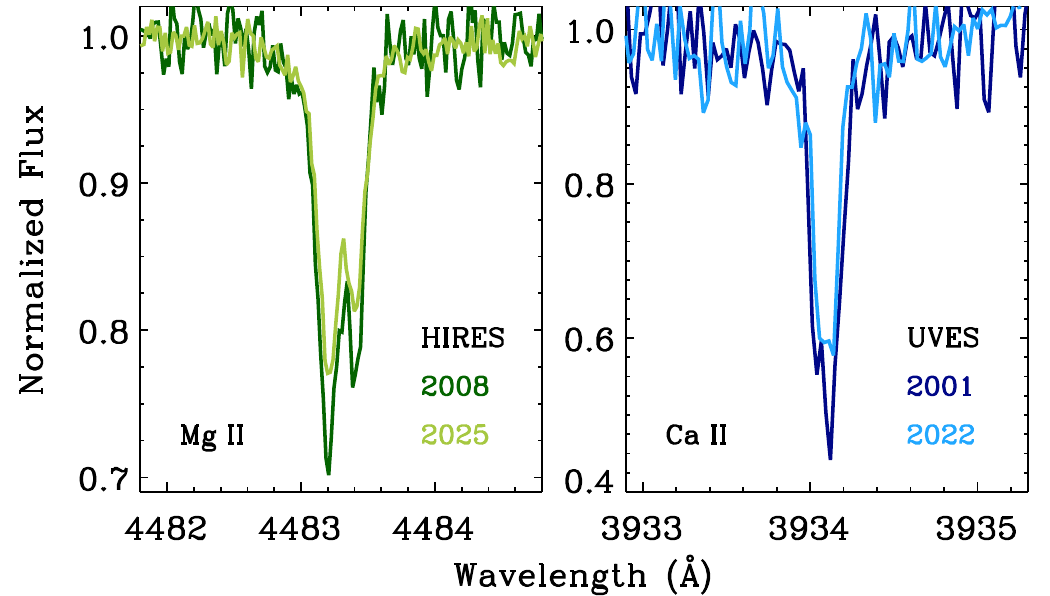}
\vskip 0pt
\caption{Equivalent width time-series measurements for \ion{Ca}{ii}\,K and \ion{Mg}{ii}\,4482\AA~in 0106$-$328.  The top panel shows the measurements for each spectral line at each epoch, and the middle panel plots their ratio.  The bottom panel shows examples of variation in each feature, seen independently in two distinct instruments and observatories.
\label{ews}}
\end{figure}

\section{Analysis and results}

Details of the ground-based observing runs are provided in Table~\ref{obs}, where both \ion{Ca}{ii}\,K and \ion{Mg}{ii}\,4482\,\AA~are detected in each spectrum, and form the basis for the time-series analysis of their equivalent widths.  The ultraviolet and optical spectroscopy are combined to provide an assessment of the bulk chemical composition of the parent body, whose debris has been previously detected via circumstellar dust and atmospheric metals.

\subsection{Line equivalent widths and changes}

Equivalent width measurements are a model-independent method for assessing line strengths and thus abundances.  The relatively narrow metal features in DAZ white dwarfs are readily isolated from the surrounding continuum and broad Balmer lines.  The data all have excellent sky subtraction and there are no significant telluric features at the relevant wavelengths.  The pixel-to-pixel sensitivity variations for all arrays have been corrected by flat fielding, and the spectral response functions of the spectrographs similarly removed by observations of standard calibration stars.  Nevertheless, systematic differences between instruments are known (e.g. velocity offsets).  Here, results are reported for changes observed independently in both UVES and HIRES, each with two epochs using the same instrument settings.

The \ion{Ca}{ii} and \ion{Mg}{ii} line equivalent widths ($W_\uplambda$) and associated errors are determined for each spectroscopic epoch using a weighted-mean combined spectrum.  After continuum normalization in the adjacent wavelength regions using a polynomial fit, the absorption features are modeled with a Gaussian profile using the \textsc{emcee} package, which implements a Markov Chain Monte Carlo algorithm \citep{foreman2013}.  The line depth, line center, and standard deviation (noise) in the data are treated as free parameters. The posterior distribution is sampled using 50 walkers evolved over 5000 steps, discarding the initial 1000 as burn-in, with flat priors adopted over plausible ranges based on the data.  The median of the posterior distribution is taken as the best-fit value, and the uncertainties are derived from the 16$^{\rm th}$ and 84$^{\rm th}$ percentiles.  

In addition to this fitting error, the following steps are taken to assess the impact of continuum normalization on the total error budget.  This uncertainty is estimated by randomly perturbing the nominal continuum level 1000 times, by drawing from within the observed scatter in continuum flux, and re-fitting the lines.  The standard deviation of the resulting distribution is adopted as the systematic error due to continuum placement, and added in quadrature with the Gaussian fitting error to yield the total uncertainty.  Based on the atmospheric modeling (Section 3.2), there is a systematic offset in the X-shooter abundance vs.\ equivalent width model grid as compared to the grids at significantly higher resolution for UVES and HIRES, and a correction is made to account for this offset.  The resulting equivalent widths and total errors are given in Table~\ref{obs}.

Figure~\ref{ews} plots the time-series measurements of $W_\ion{Ca}{ii}$, $W_\ion{Mg}{ii}$, and their ratio for the six observational epochs spanning 25\,yr.  The 2025 Oct measurements for both elements support an overall decrease of 20--30\,per cent in their equivalent widths, and corresponding roughly to a 20--60\,per cent decrease in accretion rate as compared to the average of first two UVES epochs in 2000--2001.  However, between the first and final sets of measurements, these values are not monotonically decreasing but appear to exhibit fluctuations.  The measured $W_\ion{Ca}{ii}$ values appear to be consistent within and between UVES and HIRES data, where it is consistently decreasing with the exception of the single epoch taken with X-shooter.  Ultimately, $W_\ion{Mg}{ii}$ shows an overall decrease, but with two measurements higher than the UVES discovery spectra.

\begin{table}
\begin{center}
\caption{Statistics of time-series equivalent width measurements.  The third column gives the probability that the measurements are consistent with a constant value, and the observed scatter is due to intrinsic sources of noise.}\label{pvals}
\begin{tabular}{lcc}
		
\hline

Metric					&$\upchi^2$	&$p(\upchi^2 | \upnu)$\\

\hline

$W_{\ion{Ca}{ii}}$ 	                &36.6		&$<10^{-6}$\\
$W_{\ion{Mg}{ii}}$ 	                &27.6		&$<10^{-4}$\\
$W_{\ion{Ca}{ii}}/W_{\ion{Mg}{ii}}$ &10.2		&0.07\\

\hline

\end{tabular}
\end{center}
\end{table}

Because there is no clear pattern in these two equivalent widths nor their ratio, the probability that the abundances are actually changing is evaluated statistically.  First, the $\upchi^2$ statistic is calculated for each of $W_\ion{Ca}{ii}$, $W_\ion{Mg}{ii}$, and the ratio $W_\ion{Ca}{ii}/W_\ion{Mg}{ii}$, using their total associated errors.  Second, the probability of obtaining this $\upchi^2$ value given the degrees of freedom $\upnu=5$ (for six epochs) is calculated, where the null hypothesis is that each of these metrics is constant, and the observed variations are caused by measurement uncertainty.  Table~\ref{pvals} lists the results of this exercise, where the null hypothesis can be firmly ruled out for $W_\ion{Ca}{ii}$ and $W_\ion{Mg}{ii}$ individually, with a 7\,per cent chance that $W_\ion{Ca}{ii}/W_\ion{Mg}{ii}$ is constant.

The lines \ion{Si}{ii}\,3856\,\AA~and \ion{Fe}{ii}\,5169\,\AA~are detected in both HIRES epochs, but not in any other dataset, where the 2008 and 2025 epochs yield $W_\ion{Si}{ii}=(9.4\pm4.4,9.6\pm2.6)$\,m\AA, and $W_\ion{Fe}{ii}=(23.8\pm4.4,17.0\pm1.8)$\,m\AA, respectively.  There is essentially no change in $W_\ion{Si}{ii}$, but $W_\ion{Fe}{ii}$ changes by 28.6\,per cent, which remains below $2\upsigma$ because of the errors.  In contrast, between the same two HIRES epochs, $W_\ion{Ca}{ii}$ and $W_\ion{Mg}{ii}$ change by 10.2 and 17.7\,percent, at $2.7\upsigma$ and $5.0\upsigma$ significance, respectively.  Thus while other elemental abundances may be changing, these examples and the $p(\upchi^2|\upnu)$ probabilities make it clear that the Mg and Ca abundances are variable.  It is also possible that the ratios of these four elements are changing, but this cannot be confirmed with current data.

\subsection{Stellar Parameters and Abundances}

The combined, multi-wavelength spectroscopic abundances are derived to constrain the nature of the parent body, which may shed light on the variable nature of the Mg and Ca abundances. The elements O, Si, P, S, and Fe are detected in the ultraviolet data, while Mg, Si, Ca, and Fe are observed in the multi-epoch optical spectroscopy.

The $T_{\rm eff}$ and $\log\,g$ are determined by fitting white dwarf atmospheric models \citep{koester2010,koester2020} to the flux-calibrated COS spectrum of 0106$-$328, following published methodology for similar stars with the same instrument \citep{sahu2023}. The model fluxes are first scaled using $d=68.9\pm0.2$\,pc based on the {\em Gaia} eDR3 parallax and a mass-radius relation \citep{althaus2013}, which are then fitted with the absolute fluxes to derive the best-fit parameters.  This analysis yields a hydrogen atmosphere with $T_{\rm eff}=16\,100$\,K and $\log\,g=8.01$, both within $1\upsigma$ of those derived independently from {\em Gaia} photometry alone \citep{gentile2021}.  These parameters correspond to $M_\star=0.61$\,M$_\odot$, $R_\star=0.013$\,R$_\odot$, and a cooling age of 170\,Myr.

\begin{figure}
\includegraphics[width=\columnwidth]{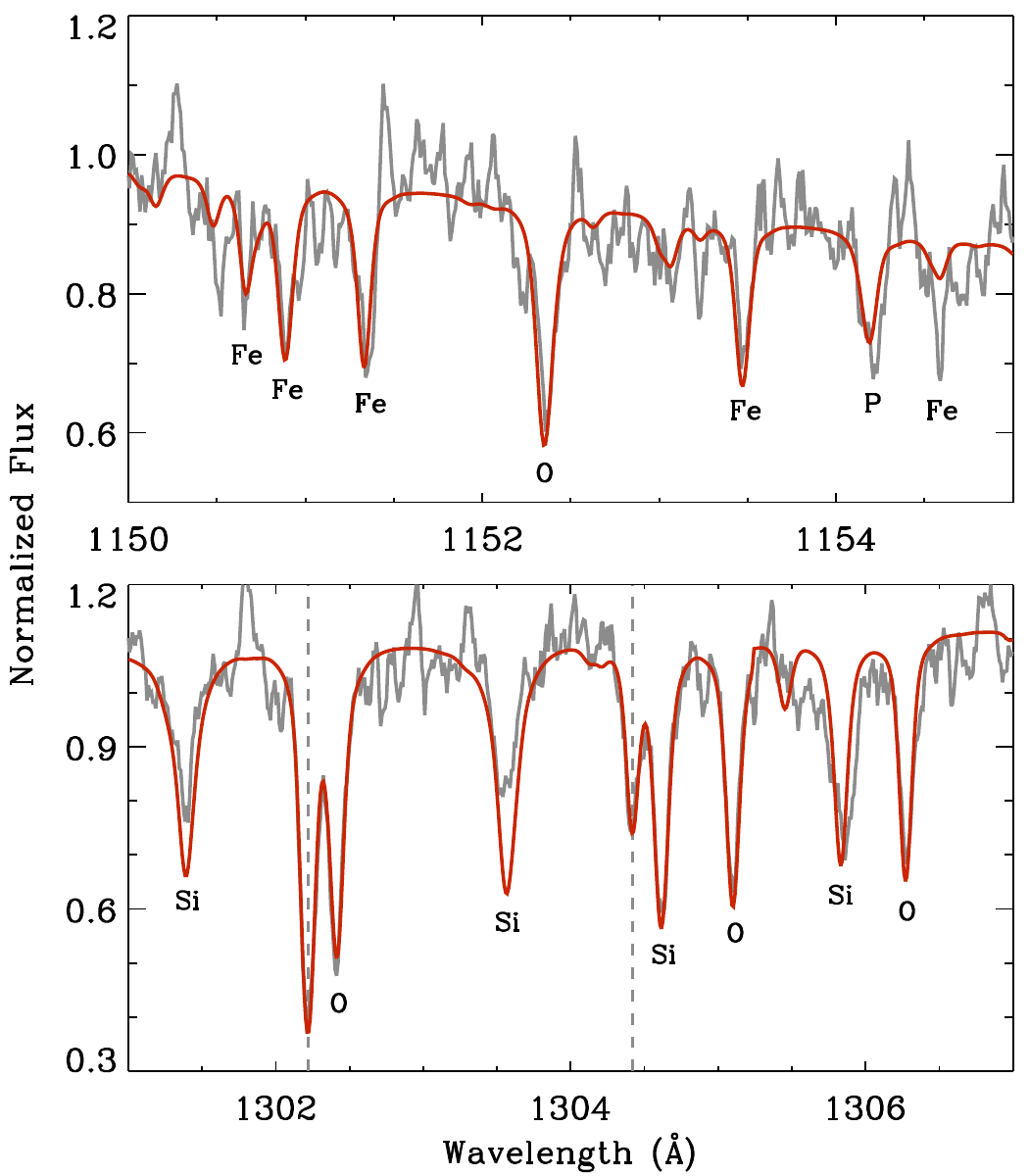}
\vskip 0pt
\caption{Two segments of the COS spectrum shown in grey, with the adopted model overplotted in red.  Absorption features are labelled by the corresponding element, and (fitted) interstellar lines are marked with dashed vertical grey lines.
\label{plotcos}}
\end{figure}

Keeping the $T_{\rm eff}$ and $\log\,g$ fixed, model atmosphere grids are generated with metal abundances in the range $-9\leq$ [Z/H] $\leq 4$, in steps of 0.5\,dex, for all relevant species based on the ultraviolet wavelength coverage. The scaled model fluxes are then fitted to the absolute fluxes, in wavelength windows centered on the photospheric absorption features, following published procedures for COS white dwarf spectra \citep{williams2025}. Because several photospheric lines are blended with interstellar features, these lines are also modeled in the fitting process. On average, the photospheric and interstellar line velocities are offset from their rest wavelengths by $52.9\pm1.4$ and $12.0\pm2.8$\,km\,s$^{-1}$, respectively. For each detected heavy element, the weighted average of the abundances derived from multiple lines is adopted as the final abundance.  Carbon is not detected and an upper limit is derived from the region corresponding to the \ion{C}{ii} doublet at 1335\,\AA. The best-fit model to the COS spectrum is shown in Figure~\ref{plotcos}.

Elemental abundances for the optical data are determined using model grids computed at the $T_{\rm eff}$ and $\log\,g$ determined above. For each element, synthetic spectra are generated for abundances $-9\leq$ [Z/H] $\leq 4$, in steps of 0.25\,dex, convolved to the appropriate instrument resolution, and the model equivalent widths are calculated for the relevant absorption lines using the method described in Section 3.1.  This procedure yields predicted line strengths as a function of abundance for each metal absorption feature and instrument, where the actual equivalent widths are interpolated onto these grids using a spline function to determine the abundance. Uncertainties in abundance are calculated by propagating the $1\upsigma$ total errors in equivalent widths and repeating the interpolation across the model line strength–abundance grid.

Although at least two metal species are varying, the changes are at most around 30--60\,per cent in mass accretion rate, and thus assessment of the bulk composition of the planetary material should be relatively robust against these modest variations.  For this purpose, the single epoch COS abundances are adopted for O, P, and S, as these elements are unique to that dataset.  The elements Mg and Ca are detected only in the optical, but their lines vary over the six ground-based spectroscopic epochs, while Si and Fe are observed both from the ground in the two HIRES epochs, and in space within the COS data.

The temporal average of the two HIRES epochs is 2017.2 and relatively close in time to the 2016.6 COS epoch.  Based on this, the logarithmic average abundances of the HIRES datasets are calculated for Mg, Si, Ca, and Fe.  From these, the resulting values for Mg and Ca are adopted for subsequent analysis, but for Si and Fe their optical means and ultraviolet measurements are combined in a logarithmic average for each.  Weighted means are taken in the optical where errors are all derived identically for single lines, but unweighted means are necessary between the optical and ultraviolet data, as it is not possible to assess line-to-line variations for a given species in the optical spectra.  The adopted abundances for the ultraviolet, optical, and combined data are summarized in Table~\ref{abunds}.

\subsection{Major rock-forming elements}

The metal-to-metal ratios of the accreted planetary material are modified by diffusion within the star, resulting in the observed photospheric abundances.  These intrinsic abundance ratios can be recovered from the atmospheric models used for line fitting and photospheric abundance determination, which encode all the relevant physics, including radiative levitation \citep[negligible for 0106$-$328;][]{koester2010,koester2020}.  The atmospheric modeling provides this information directly, under the assumption of a steady-state balance between accretion and diffusion, which results in constant diffusion flux for each element throughout the entire atmosphere.  The resulting accretion rates are given in Table~\ref{abunds}.

\begin{figure}
\includegraphics[width=\columnwidth]{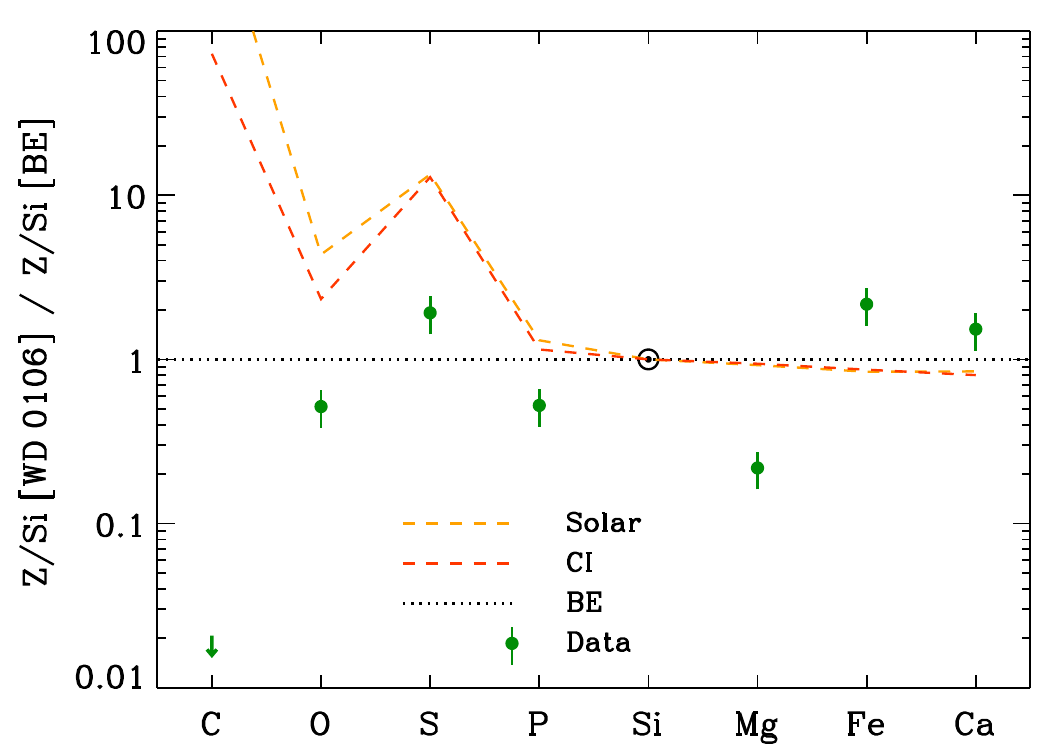}
\vskip 0pt
\caption{Abundance ratios Z/Si in the accreted material as compared to those of the bulk Earth (BE), CI chondrites, and the Sun \citep{mcdonough2000,lodders2003}, with representative error bars of 0.1\,dex shown.
\label{plotz}}
\end{figure}

The accretion rate for each heavy element is taken relative to that of Si, and these ratios Z/Si are evaluated relative to the bulk Earth, CI chondrites, and solar values.  The results for 0106$-$328 are plotted in Figure~\ref{plotz}, where the most noteworthy departures from bulk Earth and chondritic compositions are i) an extreme C depletion, ii) a notable Mg depletion, and iii) an enhancement in Fe relative to O.  These are suggestive of high temperature processing that could have removed C, and melted the parent body so that core-mantle differentiation occurred, consolidating the bulk of Fe centrally, and concentrating Mg in a mantle that was segregated prior to the ongoing accretion. 

Next, the major rock-forming elements O, Mg, Al, Si, Ca, and Fe are evaluated to constrain the overall metal oxide budget, where together these six elements constitute over 97\,per cent of the bulk Earth \citep{allegre1995}.  Because these elements combine in standard stoichiometric ratios (i.e.\ MgO, Al$_2$O$_3$, SiO$_2$, CaO), expectations can be compared with the data to better understand the accretion history \citep{klein2010}, identify excess O delivered as water in the parent body \citep{farihi2013c,raddi2015,trierweiler2025}, or infer that Fe is delivered in bulk as pure metal \citep[e.g.][]{melis2011,jura2013,williams2025}.  While there is no detection or constraining upper limit for Al, both it and Ca are lithophile elements with high condensation temperatures \citep{lodders2003}, meaning that it is essentially impossible to chemically segregate these elements and significantly alter their ratio from their cosmic or nebular abundance.  Thus, here it is assumed that Al/Ca in the material polluting 0106$-$328 is the same as their chondritic ratio.

Putting all this together, the findings are that Mg, Al, Si, and Ca have carried around 90\,per cent of the O in parent body, and thus only a small fraction of Fe is potentially present as FeO (or more rare iron oxides).  This implies that at least 60\,percent of the observed Fe originates in metallic form, and constitutes just over 40\,per cent of the total mass of the detected elements, plus Al at chondritic abundance.  These findings are consistent with a parent body that was sufficiently heated so that core-mantle differentiation could occur, resulting in a significant mass of Fe metal.  After Fe and O, the remaining mass budget of the accreting material is dominated by Si, carrying 68\,per cent of all the O.  The relative abundance of Mg is markedly low as compared to both Si and Ca, where the mass ratios Ca/Mg $=0.47$ and Mg/Si $=0.24$ most closely match that of the crust of the Earth, and are starkly distinct from those of mantle rocks \citep{rudnick2003,palme2003}.

\begin{table}
\begin{center}
\caption{Adopted abundances and calculated accretion rates for 0106$-$328.}\label{abunds}
\begin{tabular}{lrrrr}
		
\hline

Element		&Adopted	&Adopted	&Adopted	&$\dot M_{\rm z}$\\
			&COS	&HIRES	&Final	&($10^8$ g\,s$^{-1}$)\\

\hline			

C			&$<-8.6$	&		&$<-8.6$	&$<0.001$\\
O			&$-5.47$	&		&$-5.47$	&0.634\\
Mg			&		&$-5.73$	&$-5.73$	&0.091\\
Si			&$-5.65$	&$-5.25$	&$-5.45$	&0.379\\
P			&$-8.01$	&		&$-8.01$	&0.001\\
S			&$-6.90$	&		&$-6.90$	&0.025\\
Ca			&		&$-6.40$	&$-6.40$	&0.043\\
Fe			&$-6.07$	&$-5.09$	&$-5.58$	&0.820\\

\hline

$\upSigma$	&            &           &           &1.995\\

\hline

\end{tabular}
\end{center}
{\em Note.}  The COS observations detect multiple lines for each element, and the standard deviation of the fitted abundances is adopted as the error, with typical values in the range 0.1--0.2\,dex.  The HIRES abundance errors are derived from single lines and thus only the measurement errors are available, which are typically smaller than 0.05\,dex (see Table~\ref{obs} and Section 3.2).
\end{table}

\section{Discussion}

The observed changes in both the \ion{Ca}{ii} and \ion{Mg}{ii} equivalent widths in 0106$-$328 are attributed to changes in the intrinsic photospheric abundances of the corresponding elements, via steady-state accretion rate changes from its circumstellar disk.  Other possibilities can be readily dismissed.  At 16\,100\,K, the star is nowhere near the well-characterized ZZ\,Ceti instability strip for hydrogen-rich white dwarfs \citep[e.g.][]{romero2025}, and the absence of \ion{He}{i} lines in its multi-wavelength spectra demonstrates conclusively that it is hydrogen dominated.  Thus the observations cannot be from changes in surface temperature and thus opacity.  All measured line velocities are consistent with a single value and thus must be photospheric, where ground-based data yield a mean and standard deviation $55.9\pm1.0$\,km\,s$^{-1}$ for Mg, Si, Ca, and Fe \citep{xu2019b}, and the {\em HST} data yield $52.9\pm1.4$\,km\,s$^{-1}$ for multiple lines of O, Si, P, S, and Fe.

The small systematic offset in these line-of-sight velocities is almost certainly instrumental, and moreover is not statistically significant.  Circumstellar disk absorption features have been observed toward polluted white dwarfs, and exhibit velocity shifts of tens to hundreds of km\,s$^{-1}$ relative to the photosphere  \citep{gansicke2012,cauley2018,steele2021}, where the smallest expected shift would correspond to the gravitational redshift, which is around 30\,km\,s$^{-1}$ for a 0.6\,M$_\odot$ white dwarf.  Furthermore, while circumstellar \ion{Ca}{ii} absorption is plausible and has been previously observed within the disk orbiting WD\,1145+017 \citep{xu2016,redfield2017}, \ion{Mg}{ii} has a relatively high ionization energy that implies temperatures far exceeding those expected in circumstellar gas.  Therefore, variable circumstellar absorption can be dismissed.

\begin{figure}
\includegraphics[width=\columnwidth]{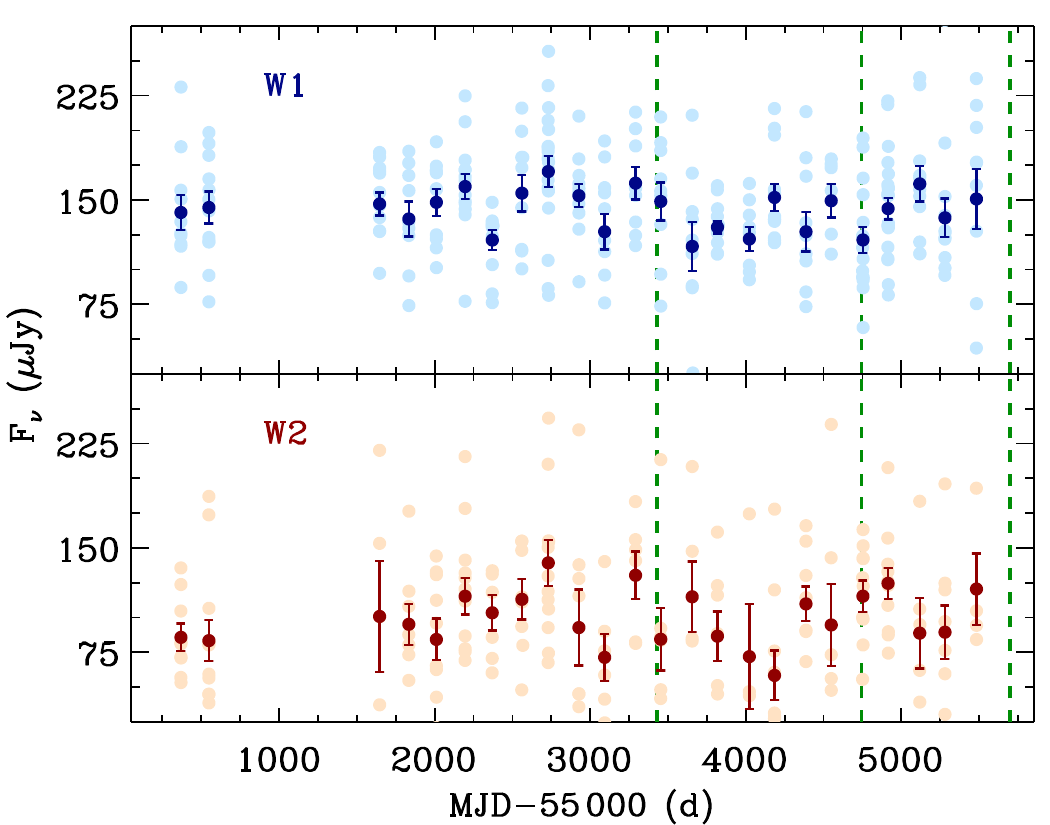}
\vskip 0pt
\caption{The infrared light curves of the debris disk orbiting 0106$-$328 as measured by {\em WISE} in channels 1 and 2.  In these plots, the starlight has been subtracted similar to published time-series dusty white dwarf studies using {\em Spitzer} \citep{swan2020,noor2025}.  The circles are individual measurements by the spacecraft and the data points with error bars are the weighted mean and standard error of the mean for each epoch of measurements.  The number of reliable detections in channel 2 and the corresponding S/N are both lower.  The green dashed vertical lines correspond to the dates of the three most recent spectroscopic epochs in Table~\ref{obs}.
\label{wise}}
\end{figure}

This star is known to have a detectable infrared excess and is thus a relatively bright outlier among the hundreds of accreting and polluted white dwarfs whose circumstellar disks lie beyond the sensitivity of {\em Spitzer} and even {\em JWST} \citep{farihi2010a,xu2012,wilsont2019,farihi2025}.  However, the dust emission is relatively subtle, with an optical depth of less than 0.1\,per cent \citep{farihi2010a,rocchetto2015}, and thus the resulting dust emission light curves as measured by {\em WISE} are modestly noisy, as can be seen in Figure~\ref{wise}.  There are adjacent epochs where the infrared disk flux changes by more than $3\upsigma$, but in general there are no clear, sustained trends in brightness, where the light curves appear stochastic, and uncorrelated with the observed changes in metal absorption lines.

\begin{figure}
\includegraphics[width=\columnwidth]{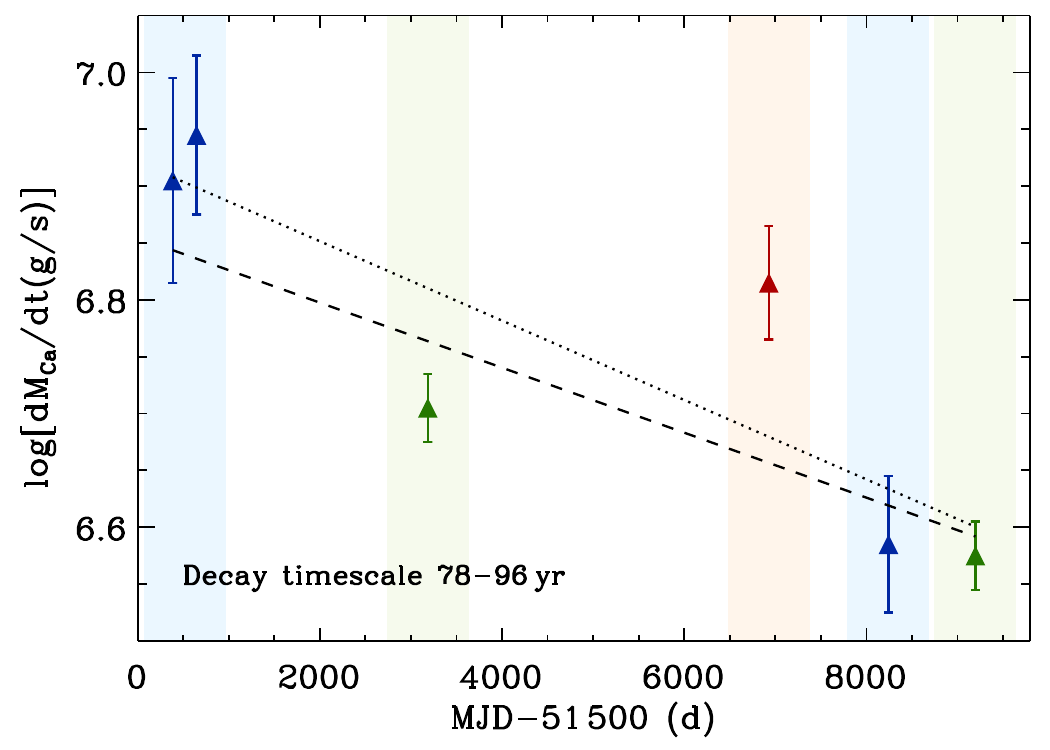}
\vskip 0pt
\caption{Same as the top panel for Figure~\ref{ews}, but now only for Ca abundances plotted as instantaneous accretion rates.  The decreasing values are fitted with a straight line whose slope is determined via $\upchi^2$ minimization, where the dashed line accounts for the individual errors, and the dotted line gives equal weight to all the data.  The slopes yield exponential decay timescales of 78 (unweighted) and 96\,yr (weighted), which compare favorably to viscous spreading in a metal-dominated, $\upalpha\approx0.1$ disk \citep{metzger2012}.
\label{mdots}}
\end{figure}

The observations of 0106$-$328 are the first compelling evidence for accretion rate changes in polluted white dwarfs.  Such changes have been predicted from two distinct sets of considerations: 1) potential short-lived bursts of high-rate accretion, inferred on the basis of differences between samples of polluted white dwarfs with relatively long vs.\ short diffusion timescales \citep{girven2012,farihi2012b}, and 2) finite accretion episodes with stochastic cadence and planetesimal mass, combined with a disk processing timescale prior to accretion; in this scenario both increases and decreases are expected \citep{wyatt2014}.  In Figure~\ref{mdots}, the relatively well-behaved \ion{Ca}{ii} abundances are transformed into a time-series of instantaneous accretion rates, and fitted with an exponential decay to constrain the disk processing timescale, which prolongs the photospheric metal lifetimes from from days to decades or longer.  The two fitted lines both use $\upchi^2$ minimization, and vary only in the weighting of each accretion rate, and yield a timescale on the order of a century, roughly speaking.

The interpretation of this timescale is not straightforward, but it is tempting to co-identify it with viscous spreading in a pure gas \citep[$\upalpha$-parameterization;][]{shakura1973} disk that sits interior to the commonly-observed disks of gas and dust detected in the optical and infrared, where all material should be sublimated prior to accretion.  However, in the model of stochastic planetesimal delivery onto white dwarfs \citep{wyatt2014}, it is expected that there is a pre-existing disc from past disruptions, and which should persist during ongoing (smaller) disruptions and would alter the simple picture of a quiescent inner gas disk. Other possibly relevant processes may include the direct injection of material on orbital timescales, or Poynting-Robertson drag, where these may not resupply the accretion disk in a simple way.  At present there are no data or theoretical considerations to distinguish the dominant disk processing mechanism, and the variations, especially in Mg, are clearly not of a simple decay.

Finally, the parent body is almost certainly differentiated, which may geochemically segregate elements at formation, but also assist in further or continued sequestration during the fragmentation process and subsequent accretion \citep[see e.g.][]{veras2017,brouwers2023}.  In the particular case of 0106$-$328, the parent material in the star appears to be a curious mixture of core and crust material, without any substantial mantle component, and this is not envisioned by any existing models \citep[e.g.][]{turner2020}.  One possibility is that the observed pollution is the result of two parent bodies, one relatively rich in a planetary crust, and the other dominated by core material.  There is at least one polluted white dwarf whose abundances are best accounted for with two parent bodies with distinct bulk chemistry \citep{johnson2022}, and that may also be the case here. The disks orbiting white dwarfs may persist for up to tens of Myr \citep{girven2012,cunningham2021}, and dynamical modeling suggests that small bodies regularly achieve star-grazing orbits on these timescales \citep{smallwood2018,li2022}.  Together, these findings suggest a pre-existing disk from a single disrupted body could be enriched by a subsequent disruption of another object.

\section{Summary}

The multi-epoch optical spectra of 0106$-$328 exhibit clear changes in the equivalent widths of both \ion{Ca}{ii} and \ion{Mg}{ii} lines.  Evaluating both of these lines individually using probability $p(\upchi^2|\upnu)$ results in a strong rejection of the null hypothesis that the observed variations are consistent with the measurement errors.  The ratio of these two equivalent widths may also be changing, and thus the ratio of these two elements within the accreted material, but there remains a 7\,per cent chance that the measured variation is due to chance.  These observations require a sufficiently short sinking timescale for heavy elements, such that any changes are observable in principle, and represent the first empirical support of metal diffusion theory in white dwarf atmospheres, following decades of theoretical underpinning \citep{fontaine1979,vauclair1979b,paquette1986,koester2009a}.

The bulk composition of the accreted planetary debris, insofar as it can be constrained given the ongoing variations, appears to be rich in Fe that was delivered in a large fraction as pure metal.  This supports mantle-core differentiation in the parent body, which would naturally segregate Fe in bulk, and where it is possible that further chemical segregation occurred by crust-mantle differentiation, and may account for the notable Mg deficit relative to Si and Ca.  Two chemically distinct parent bodies are also possible.

The behavior of the Ca equivalent widths, and the corresponding abundances and accretion rates for this element, exhibit a nearly monotonic decrease over 25\,yr.  Fitting these data with a simple linear fit in log-linear space yields a decay timescale between 78 and 96\,yr, and signals some type of disk processing.  However, it is not yet possible to attribute this to any particular physical process, but viscous disk spreading is one possibility.

It is paramount that more polluted white dwarfs with hydrogen atmospheres and short sinking timescales are observed to constrain both the magnitude and timescale of ongoing accretion variability.  This information is fundamental to this field in terms of disk formation and evolution, and also underpins the interpretation of exoplanetary bulk compositions using white dwarf metal pollution.  Regular monitoring should be carried out, at least every few years, and is critical to take the field forward.

\section*{Acknowledgements} 

The authors thank an anonymous referee for feedback that improved the manuscript.  J.~Farihi thanks J.~T.~Williams for a discussion on abundance errors, and H.~T.~Noor thanks C.~Folsom for helpful discussions on equivalent width measurement techniques.  The results presented here are based on observations made with ESO Telescopes at the La Silla Paranal Observatory, and data obtained from the ESO Science Archive Facility.  Additional observations were obtained from a NASA Keck PI Data Award, and the Keck Observatory Archive, facilitated by NASA Exoplanet Science Institute, and obtained at the W.~M.~Keck Observatory.  The authors recognize the significant cultural role and reverence that the summit of Maunakea has always had within the indigenous Hawaiian community.  Some of the research is based on observations made with the NASA / ESA Hubble Space Telescope obtained from the Space Telescope Science Institute; the observations are associated with program 14597 (PI Farihi).

\section*{Data Availability}
All data are available in the relevant observatory public archives, with the exception of the most recent 2025 HIRES observation, which is available on request.

\bigskip
\noindent
{\em Note added in proof.} During the proof stage, the authors were made aware of \ion{Mg}{ii} and \ion{Fe}{ii} observations in WD 0106$-$328 presented in the thesis of Rogers (2022). There was no variability found in \ion{Fe}{ii} but a significant offset in \ion{Mg}{ii} between two observing epochs. However, the thesis states that the evidence for variability was ``inconclusive'' (Chapter 3 pp.\,93 and 102, Chapter 5 p.\,142).

\bibliographystyle{mnras}

\bibliography{../../../references}

\bsp    
\label{lastpage}
\end{document}